\documentclass[twocolumn,showpacs,amsmath,amssymb]{revtex4}
\usepackage{graphicx}% Include figure files
\usepackage{dcolumn}% Align table columns on decimal point
\usepackage{bm}% bold math
\newcommand{\etal}{\emph{et al.}}
\newcommand{\be}{\begin{equation}}
\newcommand{\ee}{\end{equation}}
\newcommand{\bfig}{\begin{figure}}
\newcommand{\efig}{\end{figure}}

\begin{document}      

\title{Diamagnetism and Cooper pairing above $T_c$ in cuprates
} 
\author{Lu Li$^1$, Yayu Wang$^{1,2}$, Seiki Komiya$^3$, Shimpei Ono$^3$, Yoichi Ando$^{3,4}$, G. D. Gu$^5$ and N. P. Ong$^1$
}
\affiliation{
$^1$Department of Physics, Princeton University, Princeton, NJ 08544\\
$^2$Department of Physics, Tsinghua University, Beijing, China\\
$^3$Central Research Institute of Electric Power Industry, Komae, Tokyo 201-8511, Japan\\
$^4$Institute of Scientific and Industrial Research, 
Osaka University, Ibaraki, Osaka 567-0047, Japan\\
$^5$Brookhaven National Laboratories, Upton, NY 11973
}

\date{\today}      
\begin{abstract}
In the cuprate superconductors, Nernst and torque magnetization experiments 
have provided evidence that the disappearance of the Meissner 
effect at $T_c$ is caused by the
loss of long-range phase coherence, rather than the vanishing of the pair
condensate. Here we report a series of torque magnetization measurements
on single crystals of $\mathrm{La_{2-x}Sr_xCuO_4}$ (LSCO), $\mathrm{Bi_2Sr_{2-y}La_yCuO_6}$ (Bi 2201), 
$\mathrm{Bi_2Sr_2CaCu_2O_{8+\delta}}$ (Bi 2212) and optimal $\mathrm{YBa_2Cu_3O_7}$.  Some of the measurements were
taken to fields as high as 45 T.  Focusing on the magnetization above $T_c$, we show that the diamagnetic term $M_d$ appears at an onset temperature
$T^M_{onset}$ high above $T_c$.  We construct the phase diagram of both  
LSCO and Bi 2201 and show that $T^M_{onset}$ agrees with 
the onset temperature of the vortex Nernst signal $T^{\nu}_{onset}$. 
Our results provide thermodynamic evidence against a recent proposal 
that the high-temperature Nernst signal in LSCO arises from a quasiparticle 
contribution in a charge-ordered state. 
\end{abstract}


\maketitle                   % Produces the title

\section{Introduction}\label{sec:intro}
A series of experiments utilizing the Nernst Effect
\cite{XuNature00,WangPRB01,WangPRL02,WangScience03,WangPRB06}
has demonstrated that an enhanced Nernst signal is observed 
in hole-doped cuprates at temperatures
$T$ significantly above the superconducting transition 
temperature $T_c$. The high-$T$ Nernst region was identified as a continuous extension of the vortex liquid state. In this strongly fluctuating vortex-liquid state, the large Nernst signal arises from phase slippage caused by singular phase fluctuations of the pair condensate~\cite{WangPRB01,WangPRB06}.  
In the phase-disordering scenario, the unbinding of vortex-antivortex 
pairs (in zero applied $H$)
leads to the loss of long-range phase coherence at $T_c$~\cite{KivelsonNature95}.  The condensate is incapable of displaying long-range supercurrent response.
Hence, even in weak $H$, there is no Meissner effect above $T_c$
despite the survival of the pair condensate.  
Nevertheless, the persistent short-range phase stiffness supports vorticity
and produces a large, strongly $T$-dependent
Nernst signal in the presence of a temperature gradient $-\nabla T$ and 
an applied magnetic field $\bf H$. The Nernst effect above $T_c$ 
has also been investigated in Refs. \cite{BehniaPRL02,AlloulPRL06,AlloulPRL07}

Subsequently, thermodynamic evidence for the pair condensate 
above $T_c$ was obtained by torque magnetometry, which is
a very sensitive probe of diamagnetism in the cuprates~\cite{WangPRL05,LiEPL05}.
A large diamagnetic response,
that is non-linear in $H$ and grows strongly with decreasing $T$,
is specific to the Cooper pair condensate.
Consequently, the diamagnetism results present clear evidence for survival
of the pair condensate -- with sharply reduced phase stiffness -- to
temperatures high above $T_c$. This complements the transport evidence from
the Nernst experiments.
To date, the high-resolution torque measurements above $T_c$ have 
been reported in the bilayer cuprate $\mathrm{Bi_2Sr_2CaCu_2O_{8+\delta}}$ \cite{WangPRL05,LiEPL05,LiPhysica07,LiJMMM07,OngPRL07} and in lightly-doped $\mathrm{La_{2-x}Sr_xCuO_4}$ at low $T$~\cite{LiNatPhys07}. 
For earlier measurements of diamagnetism above $T_c$, see Ref. \cite{JohnstonPRB90} (on Bi 2212) and Ref. \cite{Lascialfari03} (LSCO).

We report further torque magnetization 
experiments on $\mathrm{La_{2-x}Sr_xCuO_4}$, $\mathrm{Bi_2Sr_{2-y}La_yCuO_6}$, $\mathrm{Bi_2Sr_2CaCu_2O_{8+\delta}}$ and $\mathrm{YBa_2Cu_3O_7}$, 
which show that, when a positive Nernst signal appears below the onset 
temperature $T_{onset}$, it is accompanied by a large diamagnetic
signal that grows steeply with decreasing $T$.  Extending the
torque measurements to intense fields (33 T to 45 T), we also show that 
the curves of $M$ vs. $H$ are nonlinear with a profile characteristic 
of vortex liquid response, even at elevated $T$.

Recently, the fluctuating vortex-liquid interpretation of 
the Nernst effect has been challenged by
Cyr-Choiniere \etal~\cite{LT}, who carried out Nernst measurements on a series of Nd- and Eu-doped LSCO cuprates, in which charge ordering associated with 
stripe formation is known to occur at a charge ordering temperature $T_{CO}$.  By a qualitative comparison of the Nernst coefficient 
$\nu$ in Nd-LSCO and Eu-LSCO, Cyr-Choiniere \etal ~proposed that, in pure LSCO, 
the high-temperature Nernst signal $e_N$ arises from 
small quasiparticle pockets as 
a result of Fermi-Surface (FS) rearrangement, rather than from phase-slippage 
in the pair condensate.  We discuss the problem of separating
the vortex Nernst term from quasiparticle contributions, and
the key role that diamagnetism plays in this task.  
In pure LSCO, we explain how the magnetization results reported 
here pose serious difficulties for this hypothesis.

We adopt the abbreviations LSCO 09, LSCO 12 and LSCO 17 for $\mathrm{La_{2-x}Sr_xCuO_4}$ 
with the Sr content $x$ = 0.09, 0.12 and 0.17, respectively. We use
Bi 2201 and Bi 2212 to stand for $\mathrm{Bi_2Sr_{2-y}La_yCuO_6}$ and $\mathrm{Bi_2Sr_2CaCu_2O_{8+\delta}}$, respectively, and YBCO for 
$\mathrm{YBa_2Cu_3O_7}$.
The terms underdoped, optimally-doped and overdoped are abbreviated as UD, OPT and OV, respectively.

\section{Torque Magnetometry}\label{sec:torque}
The torque magnetization was measured with the sample glued to the tip of a thin cantilever, with $\bf H$ 
applied at a tilt angle $\theta$ = 10-15$^{\rm o}$ to the crystal $c$-axis. The deflection $\varphi$ of the
cantilever by the torque was detected capacitively.  The cantilever typically 
can resolve changes in the magnetic moment of $\delta m\sim 10^{-9}$ emu. Because of the 2D electronic dispersion in cuprates, the diamagnetic orbital currents are largely confined to the $a$-$b$ plane \cite{Bergemann98,WangPRL05}. This makes torque magnetometry well-suited for detecting weak, incipient diamagnetism in cuprate crystals.  We generally report the raw data as the effective torque magnetization $M_{eff}$ defined as $M_{eff} = \tau/\mu_0HV\sin\theta$, where $\tau$ is the torque signal, $\mu_0$ the vacuum permeability, and $V$ the sample volume. ($M_{eff}$ includes all contributions to the observed torque signal.)

%%%%%%%%%%%%%%%%%%%%%%%%%%%%%%%%%%
%%%%%%%%%%%%%%%%%%%%%%%%%%%%%%%%%%
%%%%%%%%%%%%%%%%%%%%%%%%%%%%%%%%%%
%%%%%%%%%%%%%%%%%%%%%%%%%%%%%%%%%% FIGURE 1

\begin{figure}[ht]
\includegraphics[width=9.5cm]{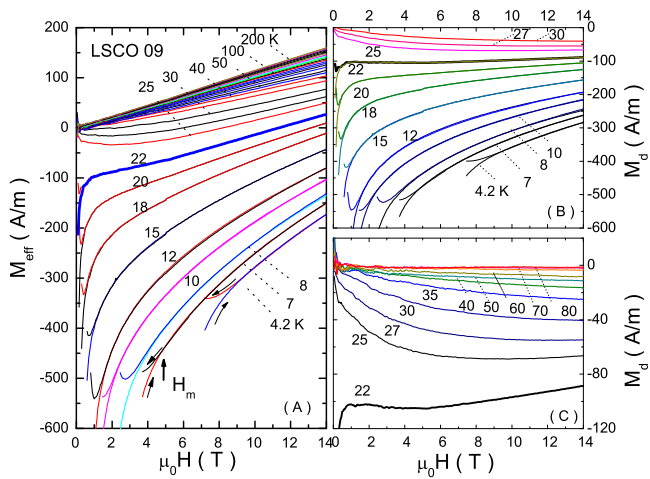}
\caption{\label{FigLSCO09}(color online)
Magnetization curves of Sample LSCO 09 with Sr content $x$ = 0.09 and transition 
temperature $T_c =$ 24 K measured in magnetic fields $H$ up to 14 T. 
(A) The (total) effective magnetization $M_{eff}$ vs. $H$ at temperatures 4.2 $\leq T \leq$ 200 K. Below $T_c$, the $M_{eff}-H$ curves are hysteretic when $H$ lies below the melting field $H_m(T)$, which is experimentally defined as
the field at which the up-sweep branch deviates from the down-sweep branch
($H_m$ is indicated by an arrow for the curve at 7 K). Above $H_m$, the curves become reversible.  (B) The diamagnetic magnetization $M_d$ vs. $H$ at temperatures 4.2 $\leq T \leq$ 30 K. (C) Curves of $M_d$ vs. $H$ at 22 K $\leq T \leq$ 80 K displayed in expanded scale.  In LSCO 09, the 
diamagnetic signal persists to more than 60 K above $T_c$. 
In (B) and (C), the bold curve is measured at the separatrix temperature $T_s$ = 22 K.
}
\end{figure}

To describe our analysis, we first examine
the curves of $M_{eff}$ vs. $H$ shown in Fig.\ref{FigLSCO09}A for Sample LSCO 09 with $T_c =$ 24 K. At temperatures $T>$ 100 K, $M_{eff}$ is strictly linear in $H$ and
paramagnetic in sign.  This reflects the dominance of the anisotropic Van Vleck paramagnetic susceptibility $\Delta\chi_{p}$, which has a weak $T$ dependence given by $\Delta\chi_{p} = A + BT$, with $A \gg BT >0$ at 200 K. (The $T$-dependence of $M_{eff}$ will be shown later in Fig. \ref{MT14T}.) The reason for measuring a dense set
of curves in the high-$T$ interval 100-200 K will emerge when we discuss extraction of
the onset temperature for diamagnetic response (Sec. \ref{sec:onset}).
Below 100 K, $M_{eff}$ begins to display a weak diamagnetic contribution that rapidly increases in magnitude as $T$ decreases. The temperature at which the diamagnetic contribution (referred to as $M_d$ hereafter) appears is identified as the onset temperature $T^M_{onset}$. 

We note that $M_{eff}$ becomes
increasingly nonlinear in $H$ as $T$ decreases from 60 K to $T_c$ (24 K).
Below $T_c$, the diamagnetic term becomes so dominant
in magnitude that $M_{eff}$ is forced to large, negative 
values despite the positive contribution of the paramagnetic Van Vleck term. 
At $T < T_c$, the $M_{eff}-H$ curves at low fields display strong hystereses between up-sweep and down-sweep branches because of strong pinning in the vortex-solid state. In the plotted curves, this is seen as a fork 
(we suppress the full hysteretic curves in both branches for clarity).
The field at which the down-sweep branch deviates from the up-sweep branch is experimentally defined as the melting field $H_m(T)$ of the vortex solid
(indicated by arrow for the curve at 7 K).  As $T$ is raised, $H_m$ decreases rapidly, reaching zero at $T_c$. A detailed investigation of $H_m(T)$
in lightly doped LSCO is reported in Ref. \cite{LiNatPhys07}.
The spin contribution to $M_{eff}$, which becomes important below 2 K in
samples with $x<$ 0.06, is also treated in detail in Ref. \cite{LiNatPhys07}.

We assume that the paramagnetic background $\Delta\chi_{p}H$ follows the trend that is seen at $T > T_{onset}$.  Hence the diamagnetic term $M_d$ is related to the observed torque magnetization $M_{eff}$ by 
\begin{equation}
M_{eff}(H) = M_d + \Delta\chi_{p}(T)H = M_d(H) + (A+BT) H.
\label{Md}
\end{equation}
Hereafter, we subtract the background Van Vleck term $(A+BT)H$ from
$M_{eff}$ and plot the diamagnetic term $M_d$ vs. $H$ (except stated otherwise).

\section{LSCO}\label{sec:lsco}
Carrying out the background subtraction for LSCO 09, we obtain the $M_d$ vs. $H$ curves. They are displayed at selected $T$ in Fig. \ref{FigLSCO09}B (4.2--30 K) and Fig. \ref{FigLSCO09}C (20--80 K). As shown in Panel B, $M_d$ is nonlinear in $H$ over a broad temperature interval. The curve at 22 K (bold curve) displays a characteristic flat profile in low fields (0.5 T $< H <$ 5 T).  We identify this temperature as
the ``separatrix'' temperature $T_s$~\cite{WangPRL05}. Below $T_s$, 
$M_d$ takes on very large, negative values at small $H$.  As $H$ increases,
$M_d$ displays an initially steep logarithmic increase, followed by a slower approach towards zero as $H$ approaches the upper critical field $H_{c2}$. The low-field curvature of the $M_d$ vs. $H$ curve changes from negative below $T_s$ to positive above $T_s$. To emphasize the high temperature diamagnetic response, panel C displays the $M_d-H$ of Sample LSCO 09 at $T \geq T_s$ in expanded scale. For $T >T_s$, the curves remain diamagnetic, displaying pronounced nonlinearity vs. $H$. We regard the nonlinear diamagnetic response above $T_c$ as clear evidence for the presence of local supercurrents as well as finite pair amplitude in the pseudogap state.  

%%%%%%%%%%%%%%%%%%%%%%%%%%%%%%%%%%
%%%%%%%%%%%%%%%%%%%%%%%%%%%%%%%%%%
%%%%%%%%%%%%%%%%%%%%%%%%%%%%%%%%%%
%%%%%%%%%%%%%%%%%%%%%%%%%%%%%%%%%% FIGURE 2
\begin{figure}[ht]
\includegraphics[width=6.5 cm]{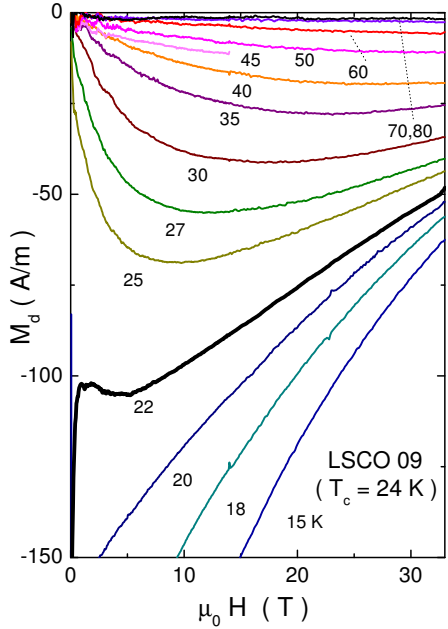}
\caption{\label{MdHLSCO09} (color online)
Magnetization curves $M_d$ vs. $H$ in LSCO 09 measured in intense fields up to 33 T.  $T_c$ lies between the curves at 22 K (bold) and 25 K.
}
\end{figure}

%%%%%%%%%%%%%%%%%%%%%%%%%%%%%%%%%%
%%%%%%%%%%%%%%%%%%%%%%%%%%%%%%%%%%
%%%%%%%%%%%%%%%%%%%%%%%%%%%%%%%%%%
%%%%%%%%%%%%%%%%%%%%%%%%%%%%%%%%%% FIGURE 3

\begin{figure}[ht]
\includegraphics[width=9.5 cm]{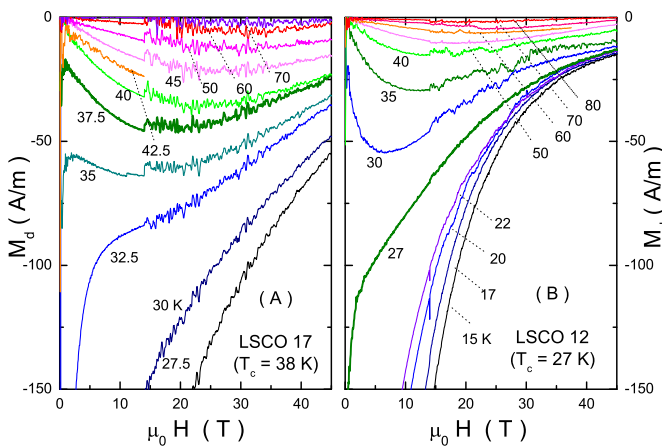}
\caption{\label{MdHLSCO2panel} (color online)
Comparison of magnetization curves $M_d$ vs. $H$ in LSC0 17 and LSCO 12 
measured in intense fields up to 45 T. Panel A shows curves for Sample LSCO 17 ($x =$ 0.17 and $T_c =$ 38 K). Panel B displays curves of Sample LSCO 12 ($x =$ 0.12 and $T_c =$ 27 K). 
}
\end{figure}

As in the superconducting state, the diamagnetic signals above $T_c$ can be suppressed by an intense magnetic field. To get a sense of how large
this field scale is, we extended the torque measurements on LSCO 09 to 33 T. After the background subtraction described above, the resulting $M_d-H$ curves are isolated and plotted in Fig. \ref{MdHLSCO09}.  For $T > T_c$, these nonlinear $M_d-H$ curves display a broad minimum. The characteristic field $H_{min}$ locating the minimum increases rapidly with $T$ (from 8 T at 25 K to 33 T at 40 K). A rough measure of the field-scale needed to observe
the inherent nonlinear response above $T_c$ is given by $H_{min}$.

We emphasize that, above 50 K, 
$M_d$ is seemingly linear in $H$.  However, this is 
simply because $H_{min}$ has now moved outside the experimental window.  
It is incorrect to attribute -- using a limited field range $H<$15 T -- this putative linear behavior to a different mechanism (e.g. quasiparticles with 
Landau diamagnetism).  Above $T_c$, $M_d$ is inherently 
non-linear in $M$ up to the onset temperature $T^M_{onset}$,
but one needs progressively higher fields to see the non-linearity
as $T$ increases.

The diamagnetic $M_d-H$ curves of Sample LSCO 17 ($x$ = 0.17, $T_c =$ 38 K) and Sample LSCO 12 ($x =$ 0.12, $T_c =$ 27 K) measured in $H$ up to 45 T are plotted in Fig. \ref{MdHLSCO2panel}A and Fig. \ref{MdHLSCO2panel}B, respectively. 
In these higher-doped crystals, the pattern of the $M_d-H$ curves 
is broadly similar to that in LSCO 09, and the foregoing
discussion applies to the diamagnetic curves. The higher field
accessed (45 T) in Fig. \ref{MdHLSCO2panel}A, B confirms the 
intrinsic non-linearity of $M_d(H)$.  In Panel A, $M_d$ is manifestly
nonlinear at all $T$ up to 70 K, even though it seems linear
when the field range is restricted to $H<$ 10 T.

The magnitude of $M_d$ in the high-field curves reveals
an interesting difference (for $T<T_c$)
between LSCO 17 and LSCO 12. The monotonic decrease of $|M_d|$
with $H$ provides an estimate of the upper 
critical field scale $H_{c2}$ (by extrapolating $M_d\to 0$).  
Comparing LSCO 17 with LSCO 12, we see that the scale of $|M_d|$ 
above 20 T is 2-3 times larger in the former at 
the same $H$ and $T$.  However,
the stronger curvature of $M_d$ vs. $H$ in LSCO 12 implies that
the decay of $M_d$ is more gradual, so that its $H_{c2}$ is 
actually higher than that in LSCO 17.  In the related cuprate
$\rm La_{2-x}Ba_xCuO_4$, stripe formation at $x=\frac18$ drives
$T_c$ to 4 K~\cite{TranquadaPRL07}. The ``dip'' in the $T_c$ dome 
in the phase diagram of LSCO suggests that fluctuating stripes 
may also exist at $x=\frac18$. If this is true, the $M_d$ results suggest that fluctuating stripes reduce the overall pair condensate
strength (compared with $x=0.17$) but allows it to survive to 
slightly larger fields.

\section{Bismuth-based cuprates}\label{sec:bi}
The nonlinear diamagnetic signals above $T_c$ are also observed in the single-layer $\mathrm{Bi_2Sr_{2-y}La_yCuO_6}$ (Bi 2201) family. In this series, the transition temperature $T_c$ is tuned by the La content $y$. The optimal $T_c$ occurs at $y \sim$0.44. Samples with $y >$ 0.44 are UD, 
while those with $y <$ 0.44 are OV.

%%%%%%%%%%%%%%%%%%%%%%%%%%%%%%%%%%
%%%%%%%%%%%%%%%%%%%%%%%%%%%%%%%%%%
%%%%%%%%%%%%%%%%%%%%%%%%%%%%%%%%%%
%%%%%%%%%%%%%%%%%%%%%%%%%%%%%%%%%% FIGURE 4

\begin{figure}[ht]
\includegraphics[width=8.5cm]{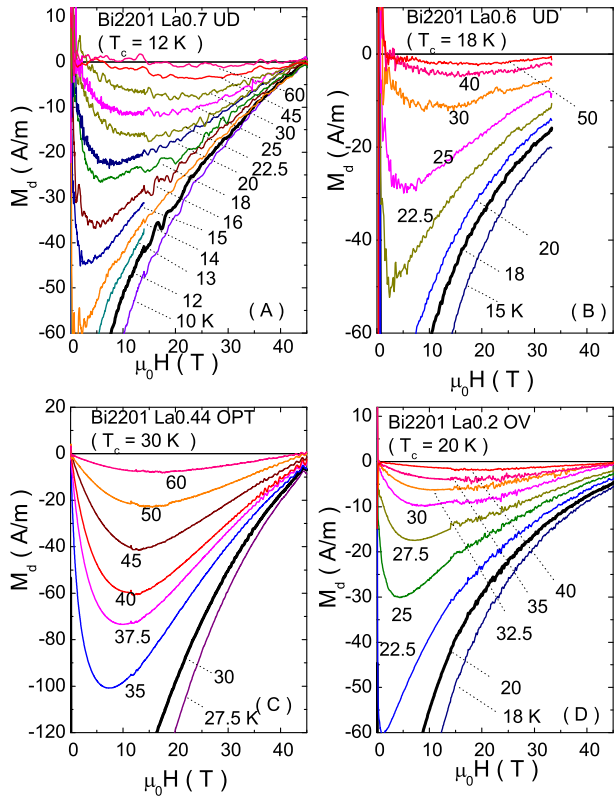}
\caption{\label{MdHBi2201} (color online)
Curves of magnetization $M_d$ vs. $H$ in Bi 2201 measured in intense fields. Panel A displays results on UD Sample Bi 2201 (La content $y =$ 0.7 and $T_c =$ 12 K).  Note that most of the results shown are at $T$ above $T_c$ = 12 K.
They approach zero at the same nominal field scale$\sim$42 T.
Panel B plots curves of $M_d$ vs. $H$ measured in another UD Bi 2201 (La content $y =$ 0.6 and $T_c =$ 18 K).
Panels C and D display $M_d$ curves measured in 
optimally-doped Bi 2201 (La content $y =$ 0.44 and $T_c =$ 30 K) and OV Bi 2201 (La content $y =$ 0.2 and $T_c =$ 20 K). 
In all these Bi 2201 samples, the curves of $M_d$ vs. $H$ remain strikingly 
nonlinear high above $T_c$.  
In each panel, the curve closest to $T_c$ is shown in bold.
A preliminary version of Panel A was published in Ref. \cite{LiPhysica07}.
}
\end{figure}

%%%%%%%%%%%%%%%%%%%%%%%%%%%%%%%%%%
%%%%%%%%%%%%%%%%%%%%%%%%%%%%%%%%%%
%%%%%%%%%%%%%%%%%%%%%%%%%%%%%%%%%%
%%%%%%%%%%%%%%%%%%%%%%%%%%%%%%%%%% FIGURE 5
\begin{figure}[ht]
\includegraphics[width=9cm]{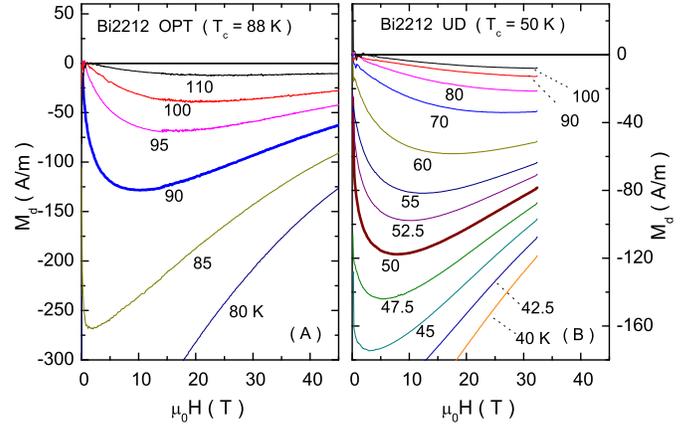}
\caption{\label{MdHBi2212} (color online)
Comparison of high-field magnetization curves $M_d$ vs. $H$ in 
OPT Bi 2212 and UD Bi 2212. Panel A \cite{LiPhysica07} displays
the $M_d$ curves in OPT Bi 2212 ($T_c =$ 88 K) in fields up to 45 T.  
Panel B plots magnetization curves in UD Bi 2212 
($T_c =$ 50 K) in fields up to 33 T.
In each panel, the curve closest to $T_c$ is shown in bold.
Panel (A) is taken from Ref.~\cite{LiPhysica07}. 
}
\end{figure}

Figure \ref{MdHBi2201} displays the diamagnetic $M_d-H$ curves of the single-layer cuprate Bi 2201 samples in the UD region (Panel A and B), in the optimally-doped region (C), and in the OV region (D). Above $T_c$, the $M_d-H$ curves in Bi 2201 are also similar to those in LSCO shown in Fig. \ref{MdHLSCO09} and Fig. \ref{MdHLSCO2panel}, except that the magnitudes of $M_d$ and field scales are slightly smaller in the former.  Above $T_c$, $M_d$ attains a broad minimum at fields below 20 T, and then approaches zero at $H \geq$ 40 T. Like the curves for LSCO in Fig. \ref{MdHLSCO09} and Fig. \ref{MdHLSCO2panel}, the curvature of the low-field $M_d-H$ curves changes from negative to positive as $T$ increases across $T_c$. As $H$ is increased beyond 20 T, $M_d$ is greatly suppressed. Above $T_c$, the complete suppression of $M_d$ requires very high fields -- comparable
to those needed below $T_c$.  Even in UD Bi 2201 with $y=$ 0.7, where $T_c=$ is quite low (12 K), the $M_d$ curves are suppressed to zero at $H \sim$ 38 - 42 T at $T \leq$ 45 K. Above $T^M_{onset}$ ($\sim$ 50 K in this sample), $M_d$ vanishes throughout our entire field range. The interesting weak-field region is 
discussed under ``fragile London rigidity''.

In Fig. \ref{MdHBi2212}, we compare the curves in OPT and UD bilayer Bi 2212
(Panels A and B, respectively). Relative to the single-layer Bi 2201, the
amplitude $|M_d|$ in Bi 2212 attains much larger values, and extend to higher field scales.  By extrapolating the low-$T$ $M_d$-$H$ curves, we estimate that $H_{c2}$ exceeds $\sim$150 T (compared with 50-80 T for Bi 2201).\\

\noindent\emph{Fragile London Rigidity}\\
One of the most interesting features of the vortex liquid state above
$T_c$ is the fragile London rigidity, observable in the
limit $H\to 0$.  In Ref. \cite{LiEPL05}, Li \etal~discovered that
over a broad interval of $T$ (86-105 K) in OPT Bi 2212, the
low-$H$ $M_d$ follows the 
power-law dependence
\be
M_d(T,H)\sim -H^{1/\delta(T)} \quad (H\to 0),
\label{power}
\ee
with an exponent $\delta(T)$ that grows rapidly from 1 (at $T\simeq$105 K)
to large values ($>$6) as $T\to T_c^{+}$.  This implies that
the weak-field diamagnetic susceptibility 
$\chi = \lim_{H\to 0} M/H \to -\infty$ is weakly divergent throughout
the interval in $T$ where $\delta>1$.  However, this divergence is
extremely sensitive to field suppression.  The 
fragile London rigidity seems to reflect the increasing tendency
of the phase-disordered condensate to establish long-range
superfluid response as $T\to T_c^{+}$.  It has no analog in bulk samples of low-$T_c$ superconductors, but may exist in a finite $T$ interval above the 
Kosterlitz-Thouless (KT) transition in 2D systems such as Mo$_{1-x}$Ge$_x$ and InO$_x$.

\begin{figure}[ht]
\includegraphics[width=8.5cm]{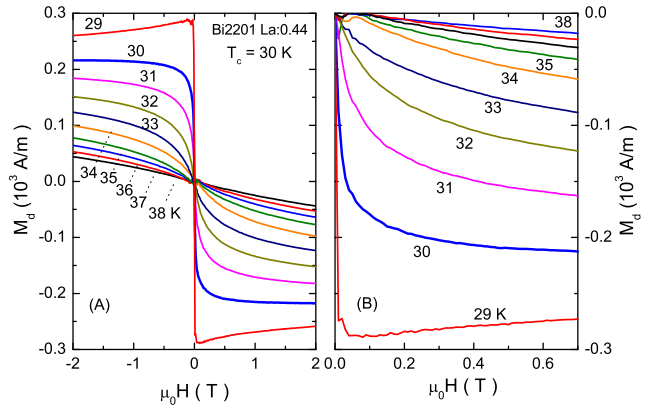}
\caption{\label{figMlow} (color online)
The low-field magnetization curves of OPT Bi2201 (La content $y =$ 0.44). 
In Panel (A), the striking weak-field non-linearity
is highlighted by displaying the variation of $M_d$ from $H$= -2 T
to +2 T.  Although $T_c$ = 30 K (bold curve), diamagnetism 
is observed up to the onset temperature $T^M_{onset}\simeq$70 K. 
Panel (B) shows the low-field curvature in expanded scale. 
The measurements were performed using a very soft 
cantilever beam. The divergent curvature of $M_d$ at 
zero $H$ is consistent with fragile London rigidity~\cite{LiEPL05,OngPRL07}.
}
\end{figure}

Using a soft cantilever, we have observed a similar pattern of
magnetization in OPT Bi 2201.  As shown in Fig. \ref{figMlow}A,
the $M_d$ curves display increasingly strong curvature 
as $H$ approaches zero from either direction.  As $T$ decreases from
38 K to $T_c$ (30 K), the zero-$H$ slope rises sharply to a 
vertical line (see expanded scale in Panel B).  The curve 
at $T_c$ (bold curve) seems to approach a logarithmic dependence vs. $H$
(equivalent to $\delta\to\infty$).  
(As may be seen by the oscillations, mechanical noise in this soft 
cantilever precludes accurate measurements for $|H|<$ 300 Oe. 
In Ref. \cite{LiEPL05}, high-resolution SQUID magnetometry 
was used to extend measurements
down to 10 Oe, but the volume of the present Bi 2201 crystal is too small for similar SQUID measurements.) Despite the lower resolution,
the divergent curvature apparent in Fig. \ref{figMlow}
is consistent with the appearance of
fragile London rigidity starting 8 K above $T_c$.  
The curves in Fig. \ref{figMlow}B are remarkably similar to those 
reported in \cite{LiEPL05,OngPRL07} for OPT Bi 2212.

The fragile London rigidity is likely to extend over a larger $T$ interval 
in UD samples.  However, it would be more difficult to disentangle the
intrinsic weak-$H$, nonlinear $M_d$-$H$ behavior from the effects of inhomogeneous broadening arising from local variations of $T_c$.  
In OPT samples,
we reason that such effects are minimized.  The bulk of the sample
has the maximum (OPT) $T_c$.  Minority regions with lower $T_c$ 
contribute only negligibly to the screening current. Hence, observation of 
the fragile state in OPT samples uncovers, in our opinion, a highly
unusual feature of cuprates that is intrinsic.\\

%%%%%%%%%%%%%%%%%%%%%%%%%%%%%%%%%%
%%%%%%%%%%%%%%%%%%%%%%%%%%%%%%%%%%
%%%%%%%%%%%%%%%%%%%%%%%%%%%%%%%%%%
%%%%%%%%%%%%%%%%%%%%%%%%%%%%%%%%%% FIGURE 6

\begin{figure}[ht]
\includegraphics[width=6.5cm]{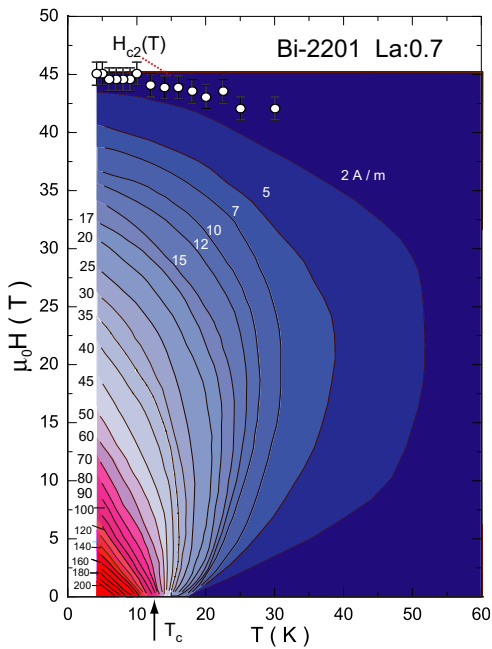}
\caption{\label{contourBi2201} (color online)
Contour plot of the diamagnetic magnetization $|M_d(T,H)|$ of UD Bi 2201,
with La content $y =$ 0.7 and $T_c =$ 12 K (arrow). 
The spacing between adjacent contour lines is 10 A/m for $T < T_c$.  
The upper critical field $H_{c2}$ (defined by extrapolating $M_d\to 0$) 
is plotted as open circles. 
}
\end{figure}

\noindent\emph{Contour Plot}\\
An instructive way to view the nonlinear diamagnetic magnetization 
is the contour plot of $M_d$ in the $T$-$H$ plane~\cite{LiJMMM07}.
Figure \ref{contourBi2201} displays the contour plot in 
single-layer UD Bi2201 (La content $y =$ 0.7, $T_c = $ 12 K). 
The value of $|M_d|$ is as indicated at selected contours.  
With $H$ fixed (e.g. at 10 T), $|M_d|$ decreases monotonically
as $T$ is raised from 4 K to 60 K.  Just as in the Nernst signal,
the diamagnetic signal in the $T$-$H$ plane 
bulges out to temperatures high above $T_c$, with
no obvious discontinuities or changes of slope.  
The highest temperature at which
$M_d$ is resolved is $\sim$50 K (the onset temperature in this
sample).  The absence of a boundary at $T_c$
implies that the vortex-liquid state below $T_c$ evolves
continuously to the diamagnetic state above $T_c$.

If we fix $T$ (e.g. at 4 K) and increase $H$, $|M_d|$ also decreases
rapidly, as plotted in Fig. \ref{MdHBi2201}. 
The field $H_{c2}$ (=45 T) at which 
$M_d\to 0$ is plotted as open circles.  With increasing $T$,
$H_{c2}(T)$ gradually decreases, roughly tracking the contour at 2 A/m.
However, unlike the mean-field BCS scenario, $H_{c2}$ remains very large at
$T_c$ (arrow).  The magnetization contour plots in Bi 2212 (see Ref. \cite{LiJMMM07}) are roughly similar to that in Fig. \ref{contourBi2201}.  
However, the scale of the magnitude $|M_d(T,H)|$ is much 
larger as expected.  $M_d(T,H)$ also extends to much higher field scales.
A particular feature is that, near $T_c$, the contours in OPT Bi 2212 are 
nearly vertical up to 33 T (the maximum applied field)
\cite{LiJMMM07}.  This implies that, at the separatrix at $T_s$, $M_d$ remains nearly $H$-independent up to 33 T.  By contrast, in Bi 2201, the
constancy extends only to 10 T, as may be seen in Fig. \ref{contourBi2201}.

%%%%%%%%%%%%%%%%%%%%%%%%%%%%%%%%%%
%%%%%%%%%%%%%%%%%%%%%%%%%%%%%%%%%%
%%%%%%%%%%%%%%%%%%%%%%%%%%%%%%%%%%
%%%%%%%%%%%%%%%%%%%%%%%%%%%%%%%%%% FIGURE 7

%%%%%%%%%%%%%%%%%%%%%%%%%%%%%%%%%%
%%%%%%%%%%%%%%%%%%%%%%%%%%%%%%%%%%
%%%%%%%%%%%%%%%%%%%%%%%%%%%%%%%%%%
%%%%%%%%%%%%%%%%%%%%%%%%%%%%%%%%%% FIGURE 8

\begin{figure}[ht]
\includegraphics[width=7cm]{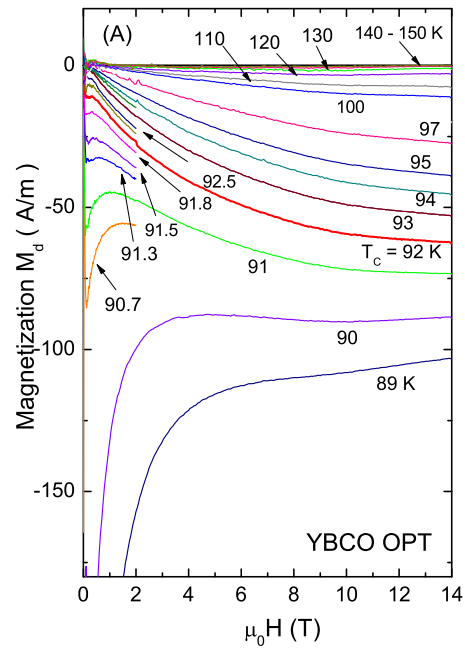}
\includegraphics[width=8cm]{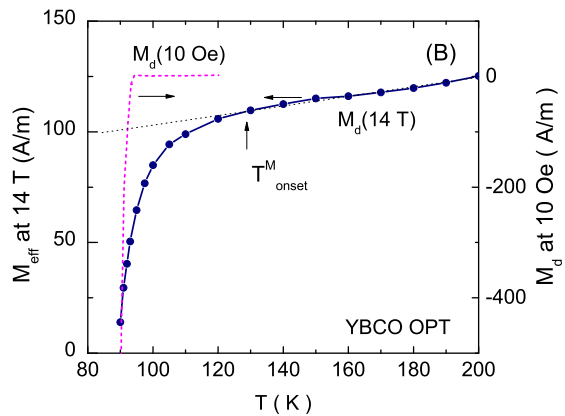}
\caption{\label{figYBCO}(color online). Magnetization curves of OPT $\rm YBa_2Cu_3O_{7-\delta}$. Panel (A) displays curves of $M_d$ vs. $H$ from $T$ = 89 
to 150 K. Some curves were taken (at slower sweep rate) only to 1 T.
Above $T_c$ = 92 K (bold curve), a sizeable diamagnetic
signal persists to $T^M_{onset}\sim$ 130 K.  Panel (B) plots the
$T$ dependence of the observed $M_{eff}$ with $H$ = 14 T. As in the other
cuprates, $M_{eff}$ may be fit to a straight line above the onset
of the diamagnetic signal $T^M_{onset}$. 
The dashed curve is the Meissner signal $M_{sq}$ measured by SQUID magnetometry
with $H$= 10 Oe. 
}
\end{figure}

\section{Optimally-doped YBCO}\label{sec:ybco}
Optimally-doped YBCO ($T_c$= 92 K) is distinguished
as the cuprate with the smallest resistivity anisotropy 
and the largest interlayer ($c$-axis) coupling energy.  
Because the coherence-length anisotropy $\xi_a/\xi_c$= 3-5
is only moderate, the vortices have the largest stiffness modulus
along $\bf c$ among cuprates ($\xi_a$ and $\xi_c$ are the coherence
lengths along the axes $\bf a$ and $\bf c$, respectively). Accordingly,
the vortex solid melting line $H_m(T)$ rises 
very rapidly below $T_c$ (to $\sim$15 T at 87 K).  
In the vortex-solid state ($H<H_m$), the dissipationless state
survives to fields of 60 T or more.
Optimally-doped YBCO should be the least susceptible to the phase
disordering mechanism for the destruction of long-range phase
coherence at $T_c$ (and hence the best candidate for 
Gaussian fluctuations among cuprates).  

However, the torque measurements reveal that $T_c$ in OPT YBCO 
is also dictated by large phase fluctuations.
Figure \ref{figYBCO}A displays the $M_d$-$H$ curves in OPT
YBCO (twinned) measured to 14 T.  The curves are broadly similar
to those in LSCO 17 and OPT Bi2212, except for the larger 
magnitude of $|M_d|$ (at comparable $H$ and $T$).
At $T_c$= 92 K, $|M_d|$ reaches the substantial value $\sim$60 A/m
at 14 T (by contrast, it should be nearly unobservable 
in a Gaussian mean-field picture).  Over the broad interval
92$\to$130 K, a large diamagnetic signal is easily observed.
As in the other hole-doped cuprates, very intense fields are needed
to suppress $M_d$ in this interval.

In Panel (B), we plot the $T$ dependence of the total torque 
magnetization $M_{eff}$ observed at 14 T (solid circles).
As in the other hole-doped cuprates, $M_{eff}$ is unresolved from the Van Vleck 
line $(A+BT)H$ until $T$ reaches $T^{M}_{onset}\sim$130 K
(arrow), below which it accelerates to very large negative values.
For comparison, we have also plotted the magnetization
measured in a very weak $H\sim$ 10 Oe (dashed curve) using
a SQUID magnetometer.  The nearly vertical decrease signals 
flux expulsion at $T_c$.  We remark
that, although $M_d$ -- measured with $H$=10 Oe -- 
is virtually unresolvable above $T_c$, the diamagnetic susceptibility 
$\chi$ is actually quite large above $T_c$ (as is clear from Panel A).
Because $M_d$ is robust to intense $H$ (100 T), 
the curve at 14 T reveals the existence of the large 
fluctuating diamagnetism associated with the vortex liquid.
This point, emphasized in Refs. \cite{WangPRL05,LiEPL05}, highlights
the major difference between the diamagnetism in 
hole-doped cuprates and low-$T_c$ superconductors. In the latter,
increasing $H$ in the fluctuation regime above $T_c$ rapidly squelches
the (Gaussian) fluctuation signal altogether. 
The curves in Fig. \ref{figYBCO}A displaying significant 
diamagnetism surviving to intense fields, at temperatures
up to 40 K \emph{above} $T_c$ is strong evidence that we are observing the phase-disordering mechanism, rather than Gaussian
mean-field fluctuations.
A comparison of the Nernst and magnetization signals in UD
YBCO is given in Ref.~\cite{Liu09}.

%%%%%%%%%%%%%%%%%%%%%%%%%%%%%%%%%%
%%%%%%%%%%%%%%%%%%%%%%%%%%%%%%%%%%
%%%%%%%%%%%%%%%%%%%%%%%%%%%%%%%%%%
%%%%%%%%%%%%%%%%%%%%%%%%%%%%%%%%%% FIGURE 9

\begin{figure}[ht]
\includegraphics[width=10cm]{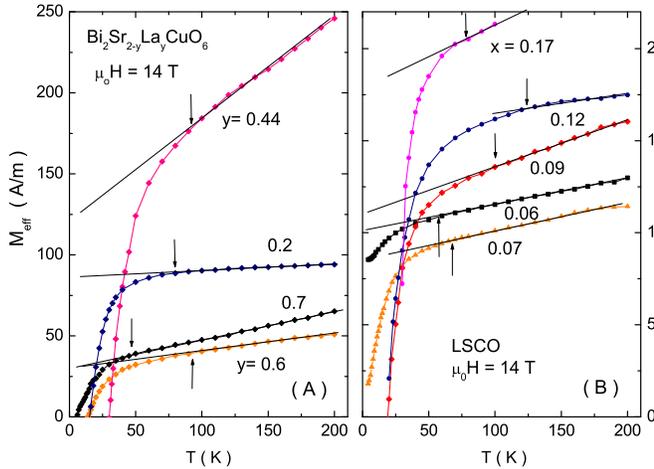}
\caption{\label{MT14T} (color online)
Plots of the temperature dependence of $M_{eff}(T)$ in Bi 2201 (Panel A)
and in LSCO (B), showing the onset of diamagnetism as $T$ is decreased.
In both panels, the value of $M_{eff}$ measured at $H$ = 14 T 
is plotted vs. $T$ in samples with various doping levels $x$. In general, 
$M_{eff}$ at high $T$ varies weakly vs. $T$, as shown by the straight lines
which are of the form $A+BT$.  Relative to this linear background,
$M_{eff}$ shows a strong downwards deviation starting
at the onset temperature $T^M_{onset}$
(indicated by arrows).}
\end{figure}

%%%%%%%%%%%%%%%%%%%%%%%%%%%%%%%%%%
%%%%%%%%%%%%%%%%%%%%%%%%%%%%%%%%%%
%%%%%%%%%%%%%%%%%%%%%%%%%%%%%%%%%%
%%%%%%%%%%%%%%%%%%%%%%%%%%%%%%%%%% FIGURE 10
\begin{figure}[ht]
\includegraphics[width=9cm]{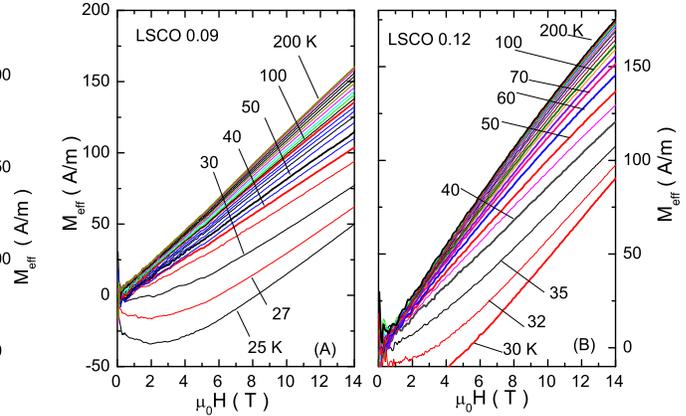}
\caption{\label{Expand} (color online)
Expanded view of the total observed magnetization $M_{eff}$ vs. $H$
in LSCO 09 (Panel A) and LSCO 12 (B). Above $T^M_{onset}$, the $T$
dependence of the curves is only from the paramagnetic
Van Vleck term.  Below $T^M_{onset}$, however,
the diamagnetic term $M_d$ grows rapidly to dominate the $T$ dependence.
The stability of the torque
cantilever and the resolution in $\tau$ are sufficient to allow closely
spaced curves to be resolved.  The uncertainty in measuring $M_{eff}$
makes the largest contribution to the error bars in $T^M_{onset}$.
}
\end{figure}
%%%%%%%%%%%%%%%%%%%%%%%%%%%%%%%%%%
%%%%%%%%%%%%%%%%%%%%%%%%%%%%%%%%%%
%%%%%%%%%%%%%%%%%%%%%%%%%%%%%%%%%%
%%%%%%%%%%%%%%%%%%%%%%%%%%%%%%%%%% FIGURE 11
\begin{figure}[ht]
\includegraphics[width=9cm]{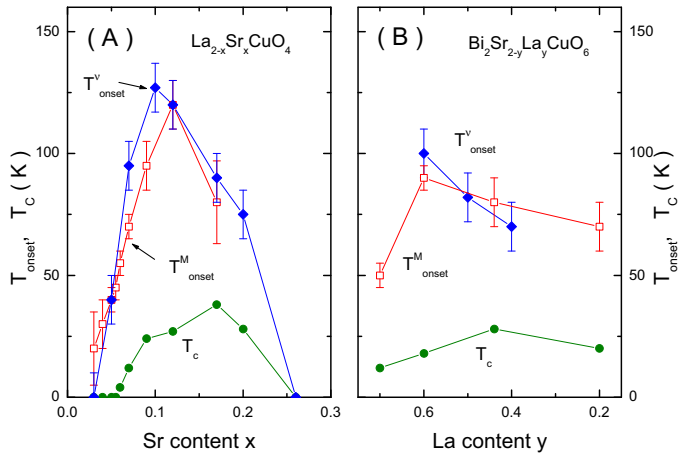}
\caption{\label{phase}(color online)
Phase diagram comparing the onset temperatures for the Nernst and diamagnetism
signals vs. doping $x$ in $\mathrm{La_{2-x}Sr_xCuO_4}$ (Panel A) and in $\mathrm{Bi_2Sr_{2-y}La_yCuO_6}$ (Panel B). 
The superconducting transition temperature $T_c$ (solid circles) is plotted with the onset temperature $T^{\nu}_{onset}$ determined by the Nernst effect (solid diamonds), and $T^M_{onset}$ determined by torque magnetometry (open squares). 
In Panel B for $\mathrm{Bi_2Sr_{2-y}La_yCuO_6}$, a large La content $y$
implies small hole carrier concentration (UD regime).
}
\end{figure}

\section{Onset temperatures and phase diagram}\label{sec:onset}
An important question is how high in temperature does the 
diamagnetic signal extend above $T_c$.  Following the procedure
in Wang \etal~\cite{WangPRL05}, we have plotted the total effective
magnetization $M_{eff}$ measured in fixed $H$ (14 T) versus $T$.
Figure \ref{MT14T} displays these plots for several samples of
Bi 2201 (Panel A) and LSCO (Panel B).  In agreement with the
results for Bi 2212 in Ref. \cite{WangPRL05}, $M_{eff}$ displays a 
weak $T$ dependence at elevated $T$ that may be fitted to the
Van Vleck anisotropy term $\Delta\chi_{p}=A+BT$ (straight lines).  The anisotropy 
$\Delta\chi_{p}$ leads to a paramagnetic torque background, as displayed.  
In each sample, $M_{eff}(T)$ displays a sharp downward deviation, 
beginning at the temperature defined as $T^M_{onset}$ (arrows).  
This reflects the appearance of local
supercurrents induced in response to the applied field.
While feeble near $T^M_{onset}$, the diamagnetic term $M_d(T)$ 
grows very rapidly in magnitude to pull $M_{eff}$ to large 
negative values.  

To determine $T^M_{onset}$ with reasonable accuracy,
it is important to establish the Van Vleck term $\chi_{p}$ with 
a sufficiently dense set of points above $T^M_{onset}$. At elevated temperatures,
the magnetization curves become closely spaced.  Thus, the stability of the
cantilever as well as the resolution in measuring $M_{eff}$ must be
sufficiently high to allow adjacent curves to be distinguished.  As examples,
we display in Fig. \ref{Expand} A and B expanded views of the curves in 
LSCO 09 and LSCO 12, respectively. The values of $M_{eff}$ at $H$ = 14 T
are the ones plotted in Fig. \ref{MT14T}.

The profile of $M_{eff}$ vs. $T$ is common to all
the samples investigated, even those in the extreme UD regime.
The characteristic profile is qualitatively different from that
seen in the gaussian regime in BCS superconductors.
Remarkably, the rapid downward acceleration of the
diamagnetic signal matches the equally rapid growth of the Nernst
signal taken at 14 T (Fig. 3 in Ref.~\cite{WangPRL05} compares
the profiles of the Nernst signal and $M_d$ measured in the 
same crystal of Bi 2212).

To compare $T^M_{onset}$ obtained here with the onset temperature
of the vortex Nernst signal $T^{\nu}_{onset}$~\cite{WangPRB01}, we plot 
the 2 onset temperatures vs. doping $x$ in the
phase diagram for LSCO (Fig. \ref{phase}A) and Bi 2201 (Fig. \ref{phase}B).
Remarkably, in LSCO, $T^M_{onset}$ (open squares) is nominally equal to 
$T^{\nu}_{onset}$ in LSCO over the entire doping range investigated.
The major difference is that the former seems to peak at $x$ = 0.12
whereas the latter peaks at 0.10.  The error bars at both temperatures
are too large to determine if the disagreement is real.  In the interesting
UD side, both temperatures decrease roughly linearly with $x$ as $x\to 0$.
[In an earlier analysis \cite{WangPRB01}, $T^{\nu}_{onset}$ was extrapolated 
below $x$ = 0.05 to reach 0 at $x=$ 0.03.
Our recent results show that this extrapolation is incorrect.  
Because the Nernst signal at $x$ = 0.03 is too weak to resolve 
even at low $T$, there is actually no experimental 
information on $T^{\nu}_{onset}$.  By contrast, the results 
on $M_{eff}$ vs. $T$ (Fig. \ref{MT14T}B) allow
$T^M_{onset}$ to be fixed reliably at small $x$.]  
Interestingly, the $x$ dependence of $H_{c2}$ obtained in Ref. \cite{LiNatPhys07} is also linear in $x$ in this regime.

In Bi 2201 (Panel B), the 2 temperature scales
are also quite similar.  However, the trend of $T^{\nu}_{onset}$ on the OV
side appears to be slightly steeper than that of $T^{M}_{onset}$.  A caveat
is that the torque measurements here were not performed on the 
same crystals as the Nernst experiments.

For the equivalent phase diagram of Bi 2212, see Ref. \cite{WangPRL05}.
The phase diagram for YBCO appears in Ref. \cite{Liu09}.

\section{Discussion}\label{sec:discuss}
\noindent\emph{Diamagnetism and Supercurrent Response}\\
The presence of a large diamagnetic response that is both
strongly $T$ dependent \emph{and} non-linear in $H$ deeply implicates
Cooper pairing.  Diamagnetism involves an orbital current density $\bf J$ that 
is antiparallel to the applied vector potential $\bf A$ 
(as in the London equation). 
Cooper pairing is -- to our knowledge -- the only established 
electronic state capable of generating the current response 
consistent with the nonlinear, strongly $T$-dependent 
diamagnetism reported here.  
[Core diamagnetism in insulators and Landau diamagnetism
(observed in pure Bi) are both
strictly $H$-linear to extremely large $H$ ($\mu_B H\simeq W$,
where $W$ is the band-width and $\mu_B$ the Bohr magneton) and nearly
$T$ independent. A ``superdiamagnetic'' state based on toroidal, orbital moments has been theorized~\cite{Ginzburg}, but this state has never been observed.] Hence, diamagnetism provides a rather direct 
detector of incipient Cooper pairing in the cuprates.

As shown in Secs. \ref{sec:lsco}, \ref{sec:bi} and \ref{sec:ybco} for LSCO,
the Bi-based cuprates and YBCO, respectively, the $M_d$-$H$ curves above $T_c$ show similar patterns - a broad minimum in moderate fields followed by a steady suppression to zero in very high fields. This pattern evolves continuously from
the curves measured below $T_c$, which display a divergence at low field caused by the Meissner effect. Above $T_c$, this divergence vanishes because of the loss of long range phase coherence.  Nonetheless, a reduced local phase rigidity survives \cite{LiEPL05}, which gives rise to the enhanced diamagnetic $M_d$ above $T_c$ at low fields. Although its overall magnitude is $\sim$10 times smaller than below $T_c$ (when observed at similar $H$), $M_d$ is readily detected as a 
strongly $T$-dependent and $H$-nonlinear response.  In UD Bi 2201, we have accomplished full field suppression of $M_d$ in fields $\sim$45 T. However, in all other cuprates, the full suppression requires fields in excess of 80 T (possibly as high as 150 T).  These impressively large
field scales are a consequence of the anomalously large binding 
energies of Cooper pairs in hole-doped cuprates.
The broad similarity of the magnetization curves in LSCO, Bi 2201, Bi 2212 and YBCO suggests that the diamagnetic behavior above $T_c$ is 
universal in the hole-doped cuprates. They are qualitatively different
from the diamagnetic response in low-$T_c$ supercondcutors.

It is instructive to compare the diamagnetism in cuprates with the
fluctuating diamagnetism observed in disordered MgB$_2$.
In a recent experiment, Bernardi \etal~\cite{Bernardi} compared the magnetization of 
pure MgB$_2$ ($T_c$ = 39 K) with disordered Mg$_{1-x}$B$_2$Al$_x$ ($x$ = 0.25, $T_c$ = 25 K).  In the disordered sample (which has a broad transition width of $\sim$15 K), 
the curves of $M_d$ vs. $H$ show that
sizeable diamagnetism exists in the narrow interval 28-32 K above its $T_c$ (Fig. 5
of \cite{Bernardi}). The profile of $M_d$ vs. $H$, which displays a 
broad minimum at $\sim$ 200 Oe, is roughly similar
to the profiles reported here (aside from the field scale). 
The broad transition width implies large inhomogneities in
the Al distribution.  As the diamagnetic response above $T_c$ does not persist
above $T_c$ of pure MgB$_2$ (39 K), we suggest that the fluctuation diamagnetism arises from
Al-poor regions of MgB$_2$ which have the highest $T_c$.  Thus, over the whole sample, diamagnetism is observable above 25 K, but not above 39 K. This
contrasts with the cuprates.  In the OPT sample within each family, 
$T^M_{onset}$ extends \emph{above} $T_c$ by factors of 1.3 (YBCO), 1.4 (Bi 2212),
2.1 (LSCO) and 2.5 (Bi 2201). Clearly, we cannot simply explain away the 
high-$T$ diamagnetism as coming from isolated OPT regions with the highest $T_c$.  
The comparison shows that the local supercurrents detected in MgB$_2$ arise from isolated regions with strong amplitude fluctuations and a broad distribution of local $T_c$'s.  By contrast, the diamagnetic signal in cuprates arises from a condensate that has lost phase stiffness, even though the gap amplitude remains large and 
nominally uniform above $T_c$.

\vspace{2mm}
\noindent\emph{Quasiparticle term in Nernst signal}\\
In UD LSCO, the quasiparticle (qp) current makes a significant contribution
to the Nernst signal $e_N$.  In the initial report
of Xu \etal~\cite{XuNature00}, the onset temperature $T_{\nu}$ was found to remain high
even when $x$ falls below 0.1 ($T_{\nu}\sim$150 K for $x$ = 0.05).  
This was traced~\cite{WangPRB01} to a significant qp contribution
to the Nernst signal.  To discuss the qp term, it is crucial to consider the
\emph{sign} of the Nernst effect.

By convention, the sign of the Nernst effect is defined as that of the
triple product ${\bf E}_N\cdot{\bf H}\times (-\nabla T)$
with ${\bf E}_N$ the observed Nernst $E$-field~\cite{WangPRB06}.
This rule is equivalent to the old convention based on
``Amperian'' current direction (clearly described by Bridgman~\cite{Bridgman}).
For vortices, ${\bf E}_N = {\bf B\times v}_L$ where 
the vortex line velocity ${\bf v}_L$
is $||(-\nabla T)$. Hence vortex flow produces a positive Nernst signal.
The qp contribution may have either sign (unrelated to their charge sign).

When the qp term is negative, it is relatively easy to separate the 
2 contributions, especially by going to intense $H$.  However, if the
qp term is positive, the separation is more difficult.
In Ref.~\cite{WangPRB01}, Wang \etal~introduced a method for separating
the qp and vortex terms by simultaneous measurements of the Hall angle $\theta_H$
and thermopower $S$ vs. $H$ to obtain the term $S\tan\theta_H$.  This subtraction procedure yields the onset temperature $T^{\nu}_{onset}$ for the vortex term
(45 K at $x$ = 0.05), which is plotted in Fig. \ref{phase}A.  With the qp
subtraction applied to the samples $x\le$ 0.07, the curve of $T^{\nu}_{onset}$
vs. $x$ has a tilted dome profile with a sharp peak at $x$ = 0.125.  For 
OPT and OV LSCO, Wang \etal~found~\cite{WangPRB06} that the qp term is \emph{negative} and negligible compared with the vortex term.  Hence, $T^{\nu}_{onset}$ gives the onset of the vortex term without the need for corrections. (The qp term is also negative
in YBCO, Bi 2201 and Bi 2212~\cite{WangPRB06}).

Recently, this assumption has been challenged by Cyr-Choiniere \etal~\cite{LT},
who proposed that rearrangement of the FS -- possibly by charge ordering or stripe formation -- produces a positive quasi-particle term that dominates the 
Nernst signal over the LSCO phase diagram at high temperatures.  
Although their Nernst measurements were largely on Eu-doped and Nd-doped LSCO
where static stripes are experimentally observed, they have extended their
hypothesis to pure LSCO.  There, they proposed that a positive qp also accounts 
for its Nernst signal at elevated $T$ (except in a narrow interval 
just above $T_c$).  

In the scenario of Ref.~\cite{LT}, the onset temperature for vortex fluctuations
should lie considerably lower than the dome of $T^{\nu}_{onset}$ plotted in Fig. \ref{phase}A.
However, the good agreement between $T^M_{onset}$ and $T^{\nu}_{onset}$ 
shows that this is not the case.  As we argued above, the strongly $T$- and $H$-dependent diamagnetism arises only from the pair condensate, and is unaffected by 
qp contributions.  The agreement between $T^M_{onset}$ and $T^{\nu}_{onset}$ 
seems to us to be strong evidence against the proposal in Ref. \cite{LT},
at least in pure LSCO.
For the claim to be viable, the hypothetical quasiparticles would have to 
produce a large, $T$-dependent diamagnetism that is also strongly non-linear
in $H$, as well as a \emph{positive} Nernst signal.

The hypothesis of Cyr-Choiniere \etal~\cite{LT} is the latest of several 
proposals (see discussion in Ref. \cite{WangPRB06}) 
that have sought to explain away the unexpected Nernst signal 
in the cuprates by invoking quasiparticles with \emph{ad hoc} properties.  
Are the large Nernst signals at high $T$ from vortices in a phase
-disordered condensate or from quasiparticles (qp)?  We argue
that torque magnetometry and the Nernst effect together constitute an incisive combination that answers this question.  
When the 2 probes show that a large positive Nernst signal coexists with a diamagnetic susceptibility (with the same onset
temperatures and similar profiles vs. $T$ and $H$), the case in favor of
phase slippage in a pair condensate with strongly disordered phase
seems compelling to us.  This is one of our main conclusions.

The magnetization-Nernst approach can also be turned around to identify
situations when the Nernst signal is not caused by vorticity.
As mentioned, in UD LSCO ($x<$0.07) the agreement of $T^M_{onset}$ and $T^{\nu}_{onset}$ provides confirmation that the subtraction proceduce based on $S\tan\theta_H$~\cite{WangPRB01} is valid. (In the broad interval
between $T^M_{onset}$ and $T_{\nu}$, extending from 45 to 150 K for $x$ = 0.05, 
there is a large positive Nernst signal, but diamagnetism is absent. Thus,
even if the qp contribution had not been identified~\cite{WangPRB01},
the present experiment would have detected the correct onset of the vortex term.)

In UD YBa$_2$Cu$_3$O$_{6+y}$ (YBCO), the qp Nernst signal 
is unusually large in a narrow window of 
doping.  In this doping range, when the qp term appears at high $T$, 
it is negative as in OV LSCO.  Significantly, the diamagnetic signal is absent
until the vortex Nernst signal appears at a lower $T$ 
(detailed YBCO results are reported in Ref.~\cite{Liu09}).
Hence, when used together, the torque magnetization and Nernst effect
readily distinguish qp from vortex contributions to the Nernst signal.  
Nernst effect and diamagnetism studies on 
$\mathrm{La_{2-x}Ba_xCuO_4}$~\cite{TranquadaPRL07} 
should provide valuable insight into the fluctuation regime in
which stripes coexist with Cooper pairing.\\

\noindent\emph{Related Experiments}\\
The surviving pair condensate above $T_c$ has also been observed in other experiments, notably in measurements of the kinetic inductance~\cite{OrensteinNature99}, STM experiments on the gap above $T_c$~\cite{YazdaniNature05}, and survival of Bogolyubov quasiparticles above $T_c$~\cite{DavisQPI}. Consistent with these observations, our results imply that the pair condensate exists well above $T_c$, surviving as
a dilute vortex liquid with local phase rigidity of short phase-correlation
length.  A number of groups recently calculated the 
Nernst signal and diamagnetism above $T_c$ in
2D superconductors and applied the results to cuprates
~\cite{MukerjeePRB04,OganesyanPRB06,
BenfattoPRL07,TesanovicNatPhy07,PodolskyPRL07}. The vortex liquid viewed as an incompressible superfluid has been treated by Anderson~\cite{AndersonNatPhy07, AndersonPRL08}. A relevant discussion
of the relation of the quantum oscillation results to the
Nernst and magnetization results is given in Ref. \cite{Lee09}.

To summarize, the high-field torque magnetometry measurements reveal that the diamagnetism persists well above $T_c$ in several families of hole-doped cuprates. As a strongly nonlinear $M_d$ vs. $H$ is 
characteristic of local supercurrent response, the diamagnetism is direct evidence that the pair condensate exists above $T_c$, surviving
in places all the way to $T^M_{onset}$.  
Hence, phase slippage is the origin of the large Nernst 
effect signals observed to that temperature.  
The agreement between the onset temperatures $T^{\nu}_{onset}$ and $T^M_{onset}$ precludes quasiparticle
interpretations for the positive Nernst signal above $T_c$.  
The magnetization results pose very serious difficulties for the quasiparticle
hypothesis~\cite{LT} proposed recently for the Nernst effect in pure LSCO.

We acknowledge numerous helpful discussions with P. W. Anderson, J. C. Davis, S. A. Kivelson, P. A. Lee, T. Senthil, Z. Te\v{s}anovi\'{c} and A. Yazdani.  The research at Princeton is supported by funds from U.S. National Science Foundation under the MRSEC Grants DMR-0213706 and DMR-0819860. Y.W. is supported by NSFC and MOST of China. G. D. G. is supported by the Department of Energy (DOE) under contract No. DE-AC02-98CH10886.  The high-field experiments were performed at the National High Magnetic Field Laboratory, which is supported by NSF Cooperative Agreement No. DMR-084173, by the State of Florida, and by the DOE.

\end{document}